\documentclass[twoside]{article}

\usepackage{PRIMEarxiv}

\usepackage[utf8]{inputenc} 
\usepackage[T1]{fontenc}    
\usepackage{hyperref}       
\usepackage{url}            
\usepackage{booktabs}       
\usepackage{amsfonts}       
\usepackage{nicefrac}       
\usepackage{microtype}      
\usepackage{lipsum}
\usepackage{fancyhdr}       
\usepackage{graphicx}       
\graphicspath{{media/}}     
\usepackage[super]{natbib}

\usepackage{graphicx}
\usepackage{amsmath,amssymb}
\usepackage{lineno}
\usepackage{multirow}
\usepackage{siunitx}
\usepackage{hyperref}
\hypersetup{
    colorlinks = true,
    allcolors = {black}
}
\usepackage{cleveref}

\crefname{equation}{Eq.}{Eqs.}
\Crefname{equation}{Equation}{Equations}

\crefname{table}{Table}{Tables}
\Crefname{table}{Table}{Tables}

\crefname{figure}{Fig.}{Figs.}
\Crefname{figure}{Figure}{Figures}

\crefname{section}{Sec.}{Secs.}
\Crefname{section}{Section}{Sections}

\crefname{chapter}{Chap.}{Chaps.}
\Crefname{section}{Section}{Sections}

\usepackage{tikz} 
\usetikzlibrary{arrows,scopes} 

\usepackage{geometry} 

\newcommand{\rc}{%
\resizebox{!}{1.25ex}{%
    \begin{tikzpicture}[>=round cap]
        \clip (0.09em,-0.05ex) rectangle (0.61em,0.81ex);
        \draw [line width=.11ex, <->, rounded corners=0.13ex] (0.1em,0.1ex) .. controls (0.24em,0.4ex) .. (0.35em,0.8ex) .. controls (0.29em,0.725ex) .. (0.25em,0.6ex) .. controls (0.7em,0.8ex) and (0.08em,-0.4ex) .. (0.55em,0.25ex);
    \end{tikzpicture}%
}%
}

\newcommand{\brc}{%
\resizebox{!}{1.3ex}{%
    \begin{tikzpicture}[>=round cap]
        \clip (0.085em,-0.1ex) rectangle (0.61em,0.875ex);
        \draw [line width=.2ex, <->, rounded corners=0.13ex] (0.1em,0.1ex) .. controls (0.24em,0.4ex) .. (0.35em,0.8ex) .. controls (0.29em,0.725ex) .. (0.25em,0.6ex) .. controls (0.7em,0.8ex) and (0.08em,-0.4ex) .. (0.55em,0.25ex);
    \end{tikzpicture}%
}%
}

\title{Efficient Aberration Correction via Optimal Bulk Speed of Sound Compensation}

\author{
  \href{https:\\www.scottschoenjr.com}{Scott Schoen~Jr}, Viksit Kumar, Yuyang Gu, \& Sunethra Dayavansha \\
  Harvard Medical School and Massachusetts General Hospital \\
  Boston, MA 02114 USA \\
  \texttt{\href{mailto:sschoenjr@mgh.harvard.edu}{sschoenjr@mgh.harvard.edu}} \\
  \AND
  Rimon Tadross \& Mike Washburn \\
  GE Healthcare \\
  Waukesha, WI 53188 \\
  \And
  Kai Thomenius \& \href{https://curt.mgh.harvard.edu/}{Anthony E. Samir} \\
  Harvard Medical School and Massachusetts General Hospital \\
  Boston, MA 02114 USA \\
}

\begin{document}
\maketitle

\begin{abstract}
Diagnostic ultrasound is a versatile and practical tool in the abdomen, and is particularly vital toward the detection and mitigation of early-stage non-alcoholic  fatty liver disease~(NAFLD).
However, its performance in those with obesity---who are at increased risk for NAFLD---is degraded due to distortions of the ultrasound as it traverses thicker, acoustically heterogeneous body walls (aberration).
Many aberration correction methods for ultrasound require measures of channel data relationships.
Simpler, bulk speed of sound optimizations based on the image itself have demonstrated empirical efficacy, but their analytical limitations have not been evaluated.
Herein, we assess analytically the bounds of a single, optimal speed of sound correction in receive beamforming to correct aberration, and improve the resulting images.
Additionally, we propose an objective metric on the post-sum B-mode image to identify this speed of sound, and validate this technique through \textit{in vitro} phantom experiments and \textit{in vivo} abdominal ultrasound data collection with physical aberrating layers.
We find that a bulk correction may approximate the aberration profile for layers of relevant thicknesses (1 to 3~cm) and speeds of sound (1400 to 1500~m/s).
Additionally, through \textit{in vitro} experiments, we show significant improvement in resolution (average point target width reduced by \SI{60}{\percent}) and improved boundary delineation \textit{in vivo} with bulk speed of sound correction determined automatically from the beamformed images.
Together, our results demonstrate the utility of simple, efficient bulk speed of sound correction to improve the quality of diagnostic liver images.
\end{abstract}

\keywords{Ultrasound \and Aberration \and NAFLD}

\section{Introduction}
\label{intro}
Diagnostic ultrasound~(US) is a versatile and practical tool in the abdomen, and has found widespread use cases including trauma assessment,\cite{rose_ultrasound_2004} vascular imaging, \cite{fadel_ultrasound_2021} and even detection of appendicitis.\cite{giljaca_diagnostic_2017} 
Of particular interest is its extension to the detection and mitigation of early-stage non-alcoholic fatty liver disease~(NAFLD), \cite{pandyarajan_screening_2019,chalasani_diagnosis_2018} which presents a large and ever-growing public health challenge.\cite{estes_modeling_2018,allen_healthcare_2018}
Ultrasound is affordable, safe, and portable compared to other modalities such as magnetic resonance imaging~(MRI) or computed tomography~(CT) imaging \cite{pandyarajan_screening_2019}, with comparable diagnostic ability (up to order \SI{85}{\percent} sensitivity and \SI{95}{\percent} specificity.\cite{hernaez_diagnostic_2011})
However, the risk of NAFLD is elevated among those with obesity \cite{cuzmar_early_2020,polyzos_obesity_2019,sarwar_obesity_2018} and these are patients for whom US imaging is degraded.\cite{de_moura_almeida_fatty_2008,mottin_role_2004}
Thus, there is a pressing need for practicable methods for improving the quality and utility of abdominal US images among this group.

Most commercial US imaging systems employ some form of conventional delay-and-sum~(DAS) algorithms, which can provide effectively real-time imaging (ca. 50 frames per second), but inherently assume that the speed of sound~(SoS) is constant throughout the medium (typically \SI{1540}{m/s} for abdominal applications).
However, as the wavefront propagates through thick subcutaneous fat layers (SoS\,$\approx$\,\SI{1450}{m/s}\cite{azhari_appendix_2010}), or indeed any heterogeneity in the speed of sound field, they are distorted such that delays appropriate for a SoS of \SI{1540}{m/s} no longer result in constructive interference of the signals, and thus the image quality is degraded.
This effect is termed ``aberration'', and is the primary source of image degradation.\cite{soulioti_deconstruction_2021-1}

Many techniques have been proposed to correct ultrasound aberration.
Some techniques correlate individual channel domain~(i.e., data from each element interpolated to each pixel position) signals with adjacent channels or a local beamsum,\cite{rigby_beamforming_1995,rigby_improved_2000} to correct for shifts introduced by an aberrating medium (termed the \emph{aberration profile}).
Other approaches model aberration as contributions from off-axis scatterers, whose influence may be compensated for to reduce their effects.\cite{krishnan_adaptive_1997,krishnan_efficient_1998} 
Recent work has proposed methods to infer aberration profiles directly from the channel domain\cite{bendjador_svd_2020} or image\cite{sharifzadeh_phase_2020} data.
Other methods look to simulate or measure an effective generated point source\cite{imbault_robust_2017,chau_locally_2019} in the medium, such that the aberration profile may be deduced and applied to each channel.
Corrections may also be computed from a known SoS distribution,\cite{ali_distributed_2018} though obtaining these distributions from ultrasound data is nontrivial.\cite{rau_ultrasound_2019,feigin_deep_2020,young_soundai_2022}
Recent image formation methods have proposed alternative beamforming techniques to suppress clutter\cite{lediju_short-lag_2011,nair_robust_2018,matrone_spatial_2021} or find optimal channel weightings\cite{ziksari_combined_2017} to mitigate incoherent contributions to the final image, though these methods often demonstrate lower inherent resolution compared to DAS in the case of low to moderate noise.\cite{lediju_bell_resolution_2015}

Additionally, a number of groups have investigated the use of a constant SoS that is different from \SI{1540}{m/s} in the beamforming to improve the resulting images, including organ-specific values \cite{barr_speed_2009}.
Anderson and Trahey\cite{anderson_direct_1998} proposed using a quadratic least squares fit to the received echoes, though this method is limited to linear arrays.
Other approaches seek the SoS that maximizes the coherence across the array,\cite{ali_local_2022} or to maximize beamformed signal's amplitude,\cite{smith_properties_1988,hyun_deep_2021} minimize its phase variation,\cite{yoon_vitro_2011} or both.\cite{perrot_so_2021} 
While such per-channel aberration correction~(AC) schemes have demonstrated improved imaging, they require analysis and subsequent manipulation of the channel domain data, which is larger than resulting image data~(i.e., the beamformer put) by a factor of $N_{e} \sim 10^{2}$.

To take advantage of the inherent efficiency of a smaller dataset, several groups have proposed metrics to characterize the speed of sound for beamforming from the B-mode images directly:
Nock et al.\cite{nock_phase_1989} proposed maximizing speckle brightness, though this technique accounts only for lateral aberration, and was observed to be less reliable for larger ROIs or noisy data.
Benjamin et al.\cite{benjamin_surgery_2018} adapted an image quality metric to characterize the SoS in homogeneous samples but did not consider effects of aberration.
Napolitano et al. proposed a spatial frequency metric to identify the optimal speed of sound from the resulting images themselves,\cite{napolitano_sound_2006} which was shown to agree with the intensity-based metric; however, this work considered only lateral spatial frequencies, and thus has inherently limited sensitivity to features that vary in the axial direction.
Yoon et al.\cite{yoon_optimal_2012} proposed a focus qualify factor based on edge conspicuity that required initial speckle reduction hat demonstrated good results \textit{in vitro}; however this approach required a finite difference approach to solve a governing diffusion equation, and clear targets to establish a whole-image gradient (that may vanish for pure speckle).
Finally, Qu et al.\cite{qu_average_2012} proposed an iterative algorithm to identify the optimal SoS from the image formed with a test SoS; however this approach requires that aberrations induce spatial shifts on the order of the speckle size, beyond which the method is unstable.
In all methods however, the effect of systematic aberration (i.e., other than random phase errors) was not evaluated.

Given the effectively real-time nature ultrasound systems, and computational constraints relative to emerging point of care devices, a simple, reliable, and efficient image-based method could provide clinical benefit.
Here, we propose such an aberration-correction scheme that, provided the receive beamforming SoS may be varied, identifies an optimal SoS for DAS beamforming from the beamformed images, we establish theoretical bounds on its applicability and effectiveness, and we show through \textit{in vitro} and \textit{in vivo} experiments that it may achieve nearly \SI{40}{\percent} improvement in resolution.


\section{Materials and Methods}
\label{sec:Methods}
The efficacy of a single, optimal SoS depends on the geometry and acoustic properties of the imaging arrangement.
Given the interest here in abdominal imaging, for which a subcutaneous fat layer is the primary source of aberration, we will consider a layered geometry and examine the theoretical validity of $c_{\mathrm{opt}}$, as well as its experimental recovery and efficacy.

\subsection*{Bulk Speed of Sound Correction Validity}
\begin{figure}[!h]
    \centering
    \includegraphics[width=0.5\textwidth]{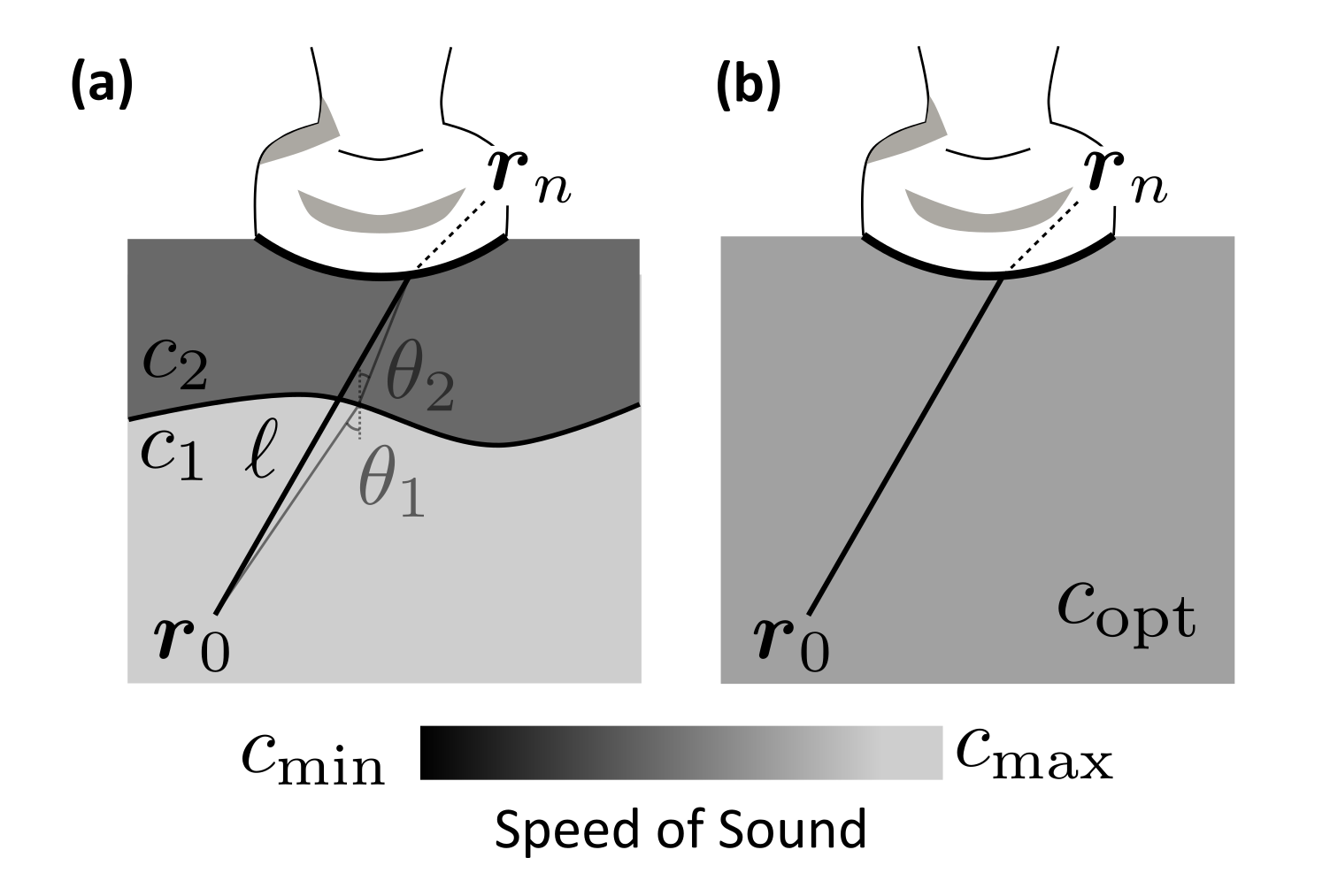}
    \caption{%
      Geometry for finding $c_{\mathrm{opt}}$. 
      \textbf{(a)}~Ground truth time-of-flight delays are computed by integrating $1/c(\boldsymbol{r})$ along the straight path $\ell$.
      \textbf{(b)}~The optimal SoS $c_{\mathrm{opt}}$ is the constant value which best approximates the delay ground truth profile.
      For convenience, the $x$-axis was taken to be tangent to the probe surface, with the origin at the center of the probe.
      }
    \label{fig:MethodsCOpt}
\end{figure}

To establish limits on the ability of a single SoS to approximate an aberration profile for a given geometry (\cref{fig:MethodsCOpt}), we first considered an analytical approach.
The appropriate receive delay corresponding to the time-of-flight between each transducer element and each point in the image is governed by propagation effects whose impact is determined by the physical properties and distribution of the medium.
This analysis contains several inherent assumptions, which are first considered.

\subsubsection*{Geometric Acoustics Assumption}
First, a geometric model of acoustic propagation (i.e., ray model) requires that the amplitude and direction of the wave vary slowly compared with the acoustic wavelength; this is reasonable for the case of abdominal ultrasound where $\lambda \sim \SI{500}{\micro\meter}$ and anatomical features of interest have scale of tens of millimeters.
Second, ultrasound in general will be refracted as it encounters an interface between two regions with different sound speeds.
For the abdominal tissues of interest, the refractive index difference has a maximal magnitude $|c_{1} - c_{2}|/c_{1 }\sim 0.1$ \cite{azhari_appendix_2010}, and thus the bending refraction may be safely neglected.
Finally, we note that the transducer has finite spatial extent in the elevation direction; 
consequently travel times may vary across the face of a single element. 
However, this variation is small (much less than the wave period for the geometries and SoS values of interest and it is a hundredfold less than the transverse variation; see \cref{sec:DelayEffects}), and thus can be excluded at a first-order approximation.

\subsubsection*{Optimal Bulk Speed of Sound}
Subject to the above considerations, we can express the true receive delay $\tau$ for each element $n$  as 
\begin{align}
\tau_{n}(\boldsymbol{r}_{0}) 
= 
\int_{\ell}{\,\frac{\mathrm{d}s}{c(\boldsymbol{r})}}\,,
\end{align}
where $\ell$ is the path between the element position $\boldsymbol{r}_{n}$ and the field point $\boldsymbol{r}_{0}$.
If we assume a uniform SoS $c_{0}$, then this integration reduces to the length of the line, and we recover an effective geometric delay $\tau^{g}$:
\begin{align}
    \tau_{n}^{g}(\boldsymbol{r}_{0})  = \frac{\| \boldsymbol{r}_{n} - \boldsymbol{r}_{0} \|}{c_{0}}\,.
\end{align}
The goodness of a bulk sound speed is quantified by the difference between the true delay and the geometric delay for a given value of $c_{0}$.
Given that a half period shift results in perfect cancellation, a quarter-wavelength receive error criterion is typically considered as the limit of coherence (i.e., the limit occurs when $c_{0}\|\tau_{n} - \tau_{n}^{g}\| \leq \lambda/4$).
We will define the optimal speed of sound $c_{\mathrm{opt}}$ for a given position in the image as that which minimizes the error across the active aperture
\begin{align}
   c_{\mathrm{opt}} = \arg\,\min_{c_{0}} \sum_{n}{\left| \tau_{n} - \tau^{g}_{n} \right|}\,.
   \label{eqn:GroundTruthOptimal}
\end{align}
Here, $n \in [1, N_{e}]$ are the indices of the subset of elements that are summed for a given $F\#$; i.e., for a point $\boldsymbol{r} = (x,z)$ the active elements are those for which $2(z-z_{n})/|x-x_{n}| > F\#$.
We note that only receive delays were assessed, as only the receive delays were modifiable in the experiments.
Thus while this analysis will hold in transmit (i.e., the errors doubled) for reciprocal transmissions (e.g., full synthetic aperture), focused transmissions would require specific analysis of the beam distortion.

In addition to evaluating the optimal SoS for a given position, we also consider the goodness of this value; that is, the value that optimizes \cref{eqn:GroundTruthOptimal} may still produce unacceptably large errors in the beamforming delays, compared to the true time of flight.
Therefore, we also consider the arrival time error $\tau_{n} - \tau^{g}_{n}$ for delays calculated with $c_{\mathrm{opt}}$ over the same region, with various layer thicknesses $d$ and SoSs $c_{\ell}$.

Finally, the mean receive delay error per element at a determined $c_{\mathrm{opt}}$ may be less than $\lambda/4$,  for relevant layer depths and thicknesses.
Therefore we evaluated the coherent aperture fraction, defined as the ratio of the largest number of contiguous elements with round trip arrival time errors less than 0.2$\lambda$ (i.e., slightly less than the destructive interference that would occur for $\lambda/4$), relative to the number of receive elements used for a given position and $F\#$.
For instance, if 96 of the 192 elements were active in receiving for a given $F\#$ and the longest subset of adjacent elements with error less than  $\lambda/5$ was 48, then the corresponding coherent aperture fraction would be 48/96 = \SI{50}{\percent}.

\subsection*{Bulk Speed of Sound Estimation}
Solving \cref{eqn:GroundTruthOptimal}  requires knowledge of the SOS distribution in the medium.
In practice however, the SoS distribution in the medium is not known, and thus the $c_{\mathrm{opt}}$ must be determined from a measure of the channel domain data itself, for example by a measure of its coherence.\cite{ali_distributed_2018}
The pixel intensity in the beamformed image is a convenient proxy for the coherence of the channel domain data, such that the latter, much larger data do not need to be manipulated, and has been shown to be reliable for material characterization.\cite{nock_phase_1989, hyun_deep_2021}

\subsubsection*{Image Metrics}
\begin{figure}
    \centering
    \includegraphics[width=0.9\textwidth]{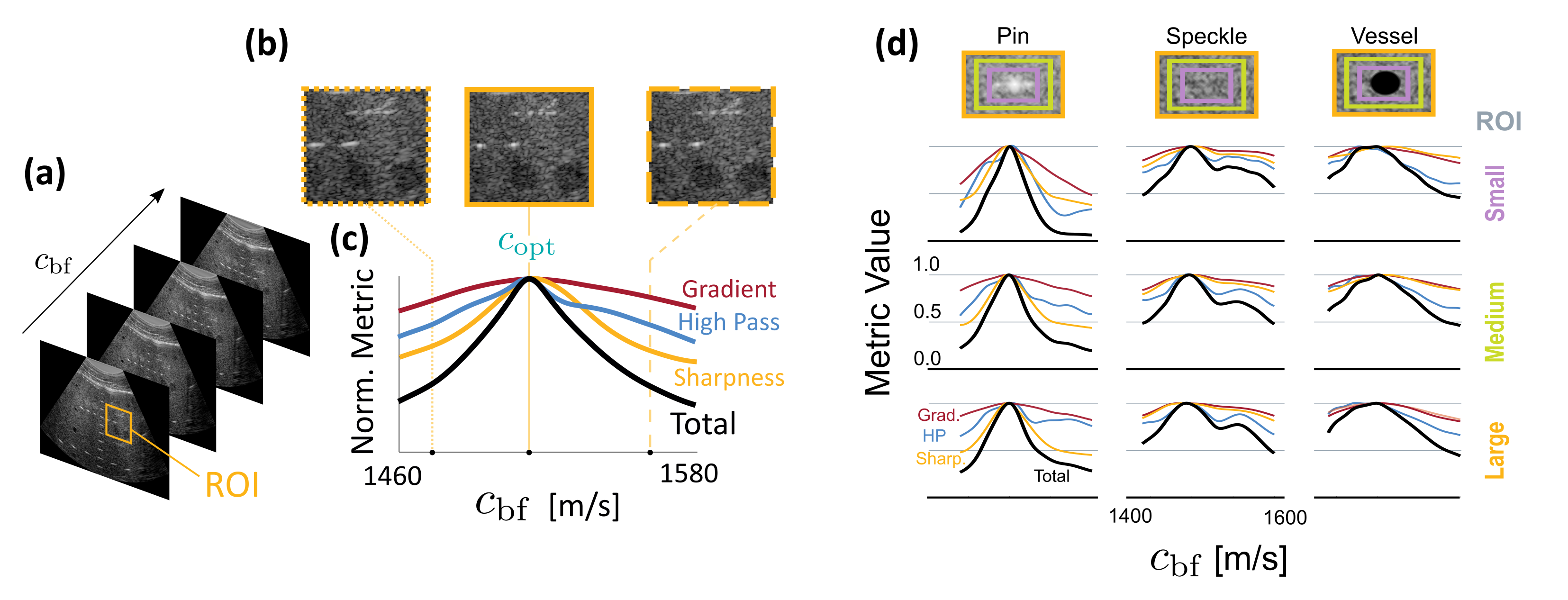}
    \caption{%
      Identification of experimental $c_{\mathrm{opt}}$ from image metrics. 
      \textbf{(a)}~Raw time series data are beamformed via conventional DAS at a range of beamforming sound speeds $c_{\mathrm{bf}}$. 
      A relevant ROI is selected.
      \textbf{(b)}~The image (pixel) data within the ROI at each $c_{\mathrm{bf}}$ are used to compute three component image metrics.
      \textbf{(c)}~The product of the three image metrics is defined as the total metric [\cref{eqn:TotalMetric}], whose maximum value is taken as the position of $c_{\mathrm{opt}}$. 
      \textbf{(d)}~Metrics are robust to target type (columns) and ROI size (rows). 
      }
    \label{fig:MethodsMetrics}
\end{figure}

We define a composite metric that is the product of three component metrics, which are computed on the DAS data prior to scan conversion and log compression for a given region of interest~(ROI) $R$, where $R \equiv \{ \boldsymbol{r}~|~(x, z) \in \text{ROI} \}$ in the image beamformed at a particular sound speed [\cref{fig:MethodsMetrics}(a--c)].
Each metric is a normalized ($M \in [0, 1]$) heuristic that measures the quality of the image region as a function of the beamforming speed of sound $c_{0}$, and their combination is employed to ameliorate any bias any might have to a particular target type.

The choice of metrics was made following experimental observations and were selected for robustness to target type and choice of ROI.
A version of the metric proposed by Ref.~\citenum{benjamin_surgery_2018} (which combines image sharpness and high spatial frequency content) was seen to provide clear maxima, but with occasional secondary peaks. 
Thus this metric was then was weighted with a gradient metric, whose value varied more slightly, but whose derivative had smaller magnitude (i.e., its maxima was clearer, but less pronounced); see \cref{fig:MethodsMetrics}(d).
The trend of a metric with $c_{\mathrm{bf}}$ was observed to be disrupted occasionally by portions of targets whose extent moved into or out of the ROI as $c_{\mathrm{bf}}$ was varied (e.g., a pin target's spread as $c_{\mathrm{bf}}$ is far from $c_{\mathrm{opt}}$); these trends were generally restored with an ROI that comprised the structures of interest for all ROIs.
Additionally, we noted that in the case of strong point scatterers, their reduced spatial extent as $c_{\mathrm{bf}} \to c_{\mathrm{opt}}$ tended to decrease the total image intensity within the ROI, and thus a brightness metric [i.e., $\sum_{R}I(x, z)$] was not used.
While pure speckle may, under random phase aberration, actually become sharper due to disadvantageous side lobes in the signal correlations, this effect is slight, and is not expected in the case of vascular or structured targets.\cite{smith_properties_1988}

First, we employ a version (see Appendix) of the image sharpness metric proposed by Refs.~\citenum{zhu_no-reference_2009,benjamin_surgery_2018}
\begin{align}
  M_{S}(R, c_{\mathrm{bf}}) = s_{1}(c_{\mathrm{bf}})\,,
\end{align}
where $s_{1}$ is the first singular value of the $N_{x}N_{z}$-by-2 matrix 
\begin{align}
    \begin{pmatrix}
      \big| & \big| \\
      \partial I/\partial x & \partial I/\partial z \\
       \big| & \big|
    \end{pmatrix}\,,
\end{align}
where the gradients are computed within the region $R$ for each $c_{\mathrm{bf}}$ (i.e., a scalar value is assigned to the ROI for each candidate SoS).
Next, the gradient metric was defined
\begin{align}
  M_{G}(R, c_{\mathrm{bf}}) 
  = 
  \sum_{R}{\Big\|\nabla I(c_{\mathrm{bf}}) \Big\|}\,,
\end{align}
where the gradient $\nabla I = ( \partial I/\partial x )\boldsymbol{e}_{x} + ( \partial I/\partial z )\boldsymbol{e}_{z}$ was computed with central difference approximations for the discrete derivatives.
Finally, a high-pass filter is applied to the data to remove lower spatial frequency components:
\begin{align}
  M_{\mathrm{HP}}(R, c_{\mathrm{bf}}) 
  = 
  \sum_{R}{ \mathcal{F}_{k}^{-1}[ \hat{I}( k \in \mathcal{K} ) ]},
\end{align}
where $\hat{I}(k_{x}, k_{z}, c_{\mathrm{bf}}) = \mathcal{F}_{k}[I(x,z,c_{\mathrm{bf}})]$ and $\mathcal{F}_{k}$ represents the 2D (spatial) Fourier transform.
Here, $\mathcal{K} = [0.75, 0.95]\,k_{\mathrm{max}}$ is the empirically selected passband of spatial frequencies, where $k_{\mathrm{max}} = \sqrt{k_{x,\mathrm{max}}^{2} + k_{z,\mathrm{max}}^{2}}$, and $k_{i,\mathrm{max}} = 2\pi/\Delta x_{i}$, with $\Delta x_{i}$ as the image pixel-spacing in dimension $i$.
The high pass metric was seen not to be very sensitive to the precise bandwidth, provided it primarily comprised spatial frequencies above $k_{\mathrm{max}}/2$ (see \cref{fig:HPFMetricBandwidths}).
Note that use of the 2D transform enables isolation of features that have sharp features in arbitrary directions, rather than the lateral direction only as analyzed in Ref.~\citenum{napolitano_sound_2006}.
The total composite metric is then defined as a product of the three component metrics
\begin{align}
  M(R, c_{\mathrm{bf}}) = M_{\mathrm{HP}} \cdot M_{G} \cdot M_{s}\,.
  \label{eqn:TotalMetric}
\end{align}
The optimal speed of sound was for the experimental data is then defined as the value of $c_{0}$ that maximizes the total metric $M$
\begin{align}
   c_{\mathrm{opt}}(R) = \arg\,\max_{c_{0}} M(R, c_{0})\,;
   \label{eqn:cOptDefinition}
\end{align}
see \cref{fig:MethodsMetrics}.
To preserve computational efficiency, images were beamformed by setting the receive beamforming SoS with step size $\Delta c_{0} = \SI{10}{m/s}$.
The image stack was then spline interpolated to \SI{1}{m/s}, as the pixel intensity varied smoothly. 
To preserve scaling between the images, the range mapping was scaled according to $\Delta r \rightarrow \Delta r(c_{\mathrm{bf}}/c_{\mathrm{ref}})$, where origin is at the center of the face of the curved probe, with the $x$-axis tangent to it.
For comparison, we all also computed the metrics proposed by Refs.~\citenum{yoon_optimal_2012} and \citenum{qu_average_2012}, with the latter metric re-scaled such that $M_{\mathrm{Qu}}' = 2 - M_{\mathrm{Qu}}/\operatorname{min}{M_{\mathrm{Qu}}}$, such that like the other metrics $M_{\mathrm{Qu}}' \in [0, 1]$ and its maximum value is sought.

\subsection*{\textit{In Vitro}, \textit{Ex Vivo}, and \textit{In Vivo} Experiments}
For the experimental data, harmonic channel data were acquired with a commercial ultrasound system (GE LOGIQ E10, GE Healthcare, Chicago, IL, USA) and a tissue-mimicking ultrasound phantom (Model 040GSE, CIRS, Norfolk, VA USA), using a curvilinear abdominal probe (GE C1-6-D, \SI{3.5}{MHz} center frequency, 192 elements, \SI{350}{\micro\meter} pitch, \SI{66}{mm} elevation focus) with default abdominal settings (\SI{3.0}{MHz}).
Image data (i.e., beamformed IQ data, from $N = 210$ scan lines) were formed via conventional delay-and-sum with a hard-coded aperture, and with receive delays from the user-specified speed of sound $c_{\mathrm{bf}}$ (i.e., transmit delays were computed using \SI{1540}{m/s}, as was used during acquisition).
The final image stack $I(\boldsymbol{r}, c_{\mathrm{bf}})$ represents the pre-scan-converted images for each bulk SoS.

Aberration was induced using material layers positioned between the surface of the phantom and the probe.
The layers included: a gel layer \SI{3}{cm} thick, with scatters (custom, \SI{0.5}{\percent} volume fraction  \SI{50}{\micro\meter} glass beads, gel speed of sound approximately \SI{1450}{m/s}, Sun Nuclear, Melbourne, FL, USA), as well as \textit{ex vivo} porcine fat samples obtained from supermarket butchers, selected visually for fat and fascia distribution.
To quantify resolution and contrast \textit{in vitro}, 11 targets (six point targets and five contrast targets, \SI{1}{cm} to \SI{10}{cm} below the fat layer) were demarcated, and the $c_{\mathrm{opt}}$ values were determined from \cref{eqn:cOptDefinition}.
The animal fat layers were heated in a water bath to approximately \SI{37}{\degree C} to best approximate realistic \textit{in vivo} material properties.
Finally, \textit{in vivo} images of the gallbladder were collected from healthy volunteers following verbal consent ($N=5$, MGH IRB approval 2022P000850) with (for those with small skin-to-capsule distances) and without (for those with more inherent aberration) the gel aberrator.

\section{Results}
\label{sec:Results}

\subsection*{Effect of Geometry on Optimal SoS}
\begin{figure}[!h]
    \centering
    \includegraphics[width=0.35\textwidth]{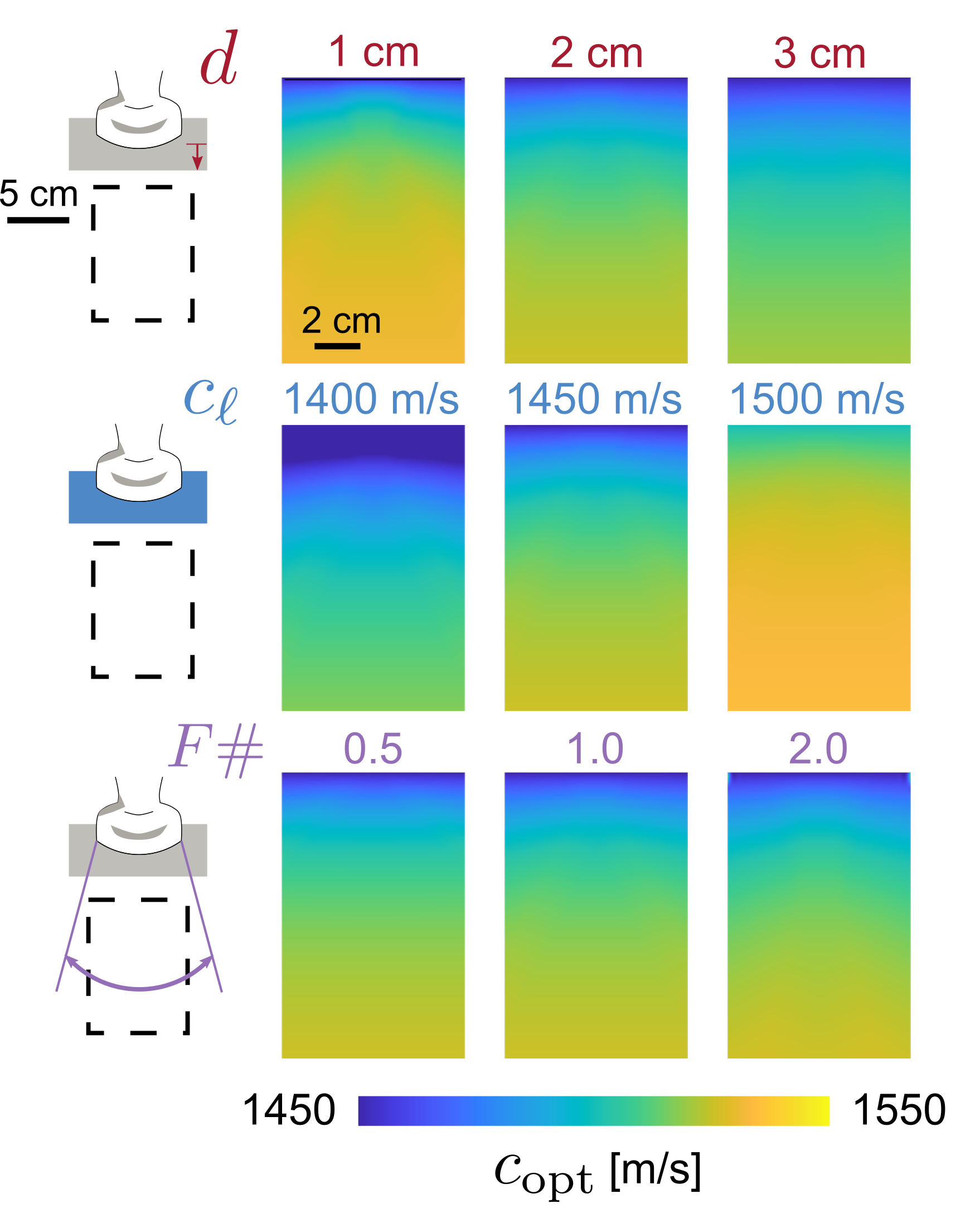}
    \caption{%
      Spatial variation in the optimal beamforming SoS $c_{\mathrm{opt}}$ [calculated from \cref{eqn:GroundTruthOptimal}].
      Values calculated for the region indicated at left, and for the indicated layer thicknesses, layer SoS $c_{\ell}$, and receiving aperture for the indicated $F\#$.
      The center column represents the default values for each parameter.
      }
    \label{fig:ResultsSoS}
\end{figure}

The optimal SoS was determined from \cref{eqn:GroundTruthOptimal} for a typical imaging field that was \SI{8}{cm} laterally by \SI{12.5}{cm} axially in a medium with SoS of \SI{1540}{m/s}, immediately beneath a layer with SoS $c_{\ell}$, thickness $d$, and with an aperture governed by $F\#$; see \cref{fig:ResultsSoS}.
For a constant layer SoS $c_{\ell} = \SI{1450}{m/s}$ (top row), thinner layers introduced larger spatial gradient in $c_{\mathrm{opt}}$ (i.e., the value of $c_{\mathrm{opt}}$ varied more rapidly in space). 
However in these cases the value of $c_{\mathrm{opt}}$ was on average closer to \SI{1540}{m/s}.
For a constant layer thickness of $d = \SI{2}{cm}$ (middle row), the spatial distribution of $c_{\mathrm{opt}}$ was similar for various $c_{\ell}$, however slower layer SoS lowered $c_{\mathrm{opt}}$, especially immediately below the layer.
When both $c_{\ell}$ and $d$ were fixed, the aperture size (defined from the $F\#$ did not have a large influence on the distribution or value of $c_{\mathrm{opt}}$ (bottom row).

\begin{figure}
    \centering
    \includegraphics[width=0.5\textwidth]{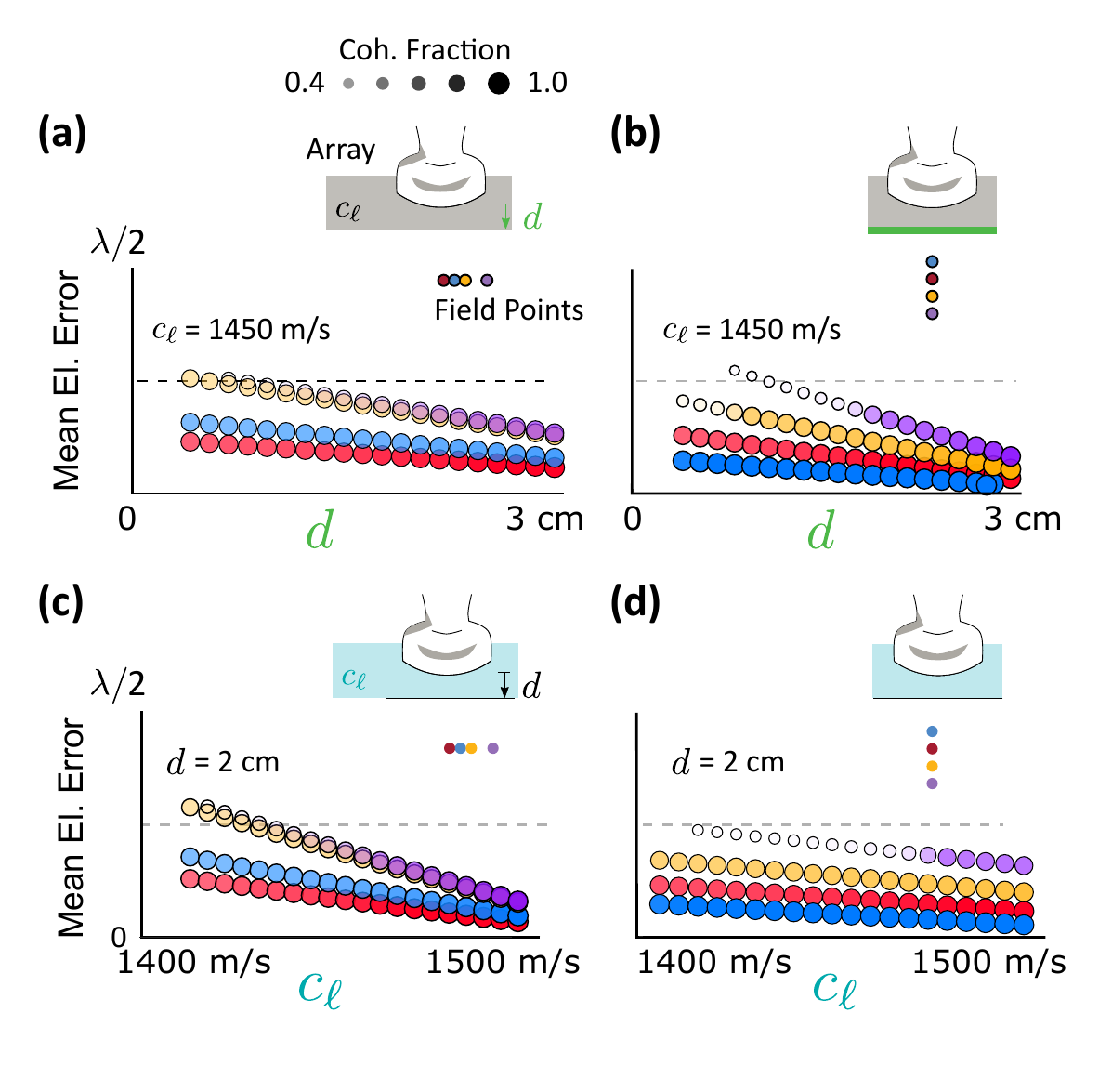}
    \caption{%
      Mean per-element arrival time error magnitude compared to ground truth (i.e., $\left\| \tau_{n} - \tau^{g}_{n}\right\|$, where geometric delays are computed with $c_{\mathrm{opt}}$) at the indicated positions.
      The size and transparency indicates the coherent aperture fraction at the given $c_{\mathrm{opt}}$.
      Reported in wavelengths $\lambda$ for \SI{3}{MHz} (\SI{333}{ns}), and for $F\# = 1.41$.
      \textbf{(a)}~Mean per-element error for a layer with $c_{\ell} = \SI{1450}{m/s}$ at the indicated axial and \textbf{(b)}~transverse positions as a function of the layer thickness $d$.
      \textbf{(c)}~Mean per-element error for a layer with $d = \SI{2}{cm}$ at the indicated axial and \textbf{(d)}~transverse positions as a function of the layer SoS $c_{\ell}$.
      }
    \label{fig:ResultsError}
\end{figure}

While the value of $c_{\mathrm{opt}}$ varied relatively smoothly with position, \cref{fig:ResultsError} illustrates the effectiveness of this SoS to keep the mean per-element error sufficiently small.
Interestingly, the mean arrival time error was largest for thinner layers, and decreases with $d$ [\cref{fig:ResultsError}(a--b)].
This is likely because thinner layers introduce more rapid spatial variation in the aberration profile across the face of the transducer, which cannot be fully accounted for by any $c_{\mathrm{opt}}$.
However, we note that the fraction of the aperture (see definition below) that could be made coherent with a single $c_{\mathrm{opt}}$ increased with $d$, as the point-to-point variation in optimal SoS varies less with thicker layers (\cref{fig:ResultsSoS}).
The influence of layer SoS for a fixed layer thickness $d = \SI{2}{cm}$, also decreased monotonically with $c_{\ell}$, expectantly vanishing as $c_{\ell}$ approaches the SoS in the underlying medium [\cref{fig:ResultsError}(c--d)].
In all cases, we find an optimal SoS may achieve acceptably small mean errors (less than $\lambda/4$ at \SI{3}{MHz} or \SI{83}{ns}) for uniform layers with realistic thicknesses and speeds of sound, and performance is generally improved on-axis and for faster, thicker layers.  

\begin{figure}
    \centering
    \includegraphics[width=0.49\textwidth]{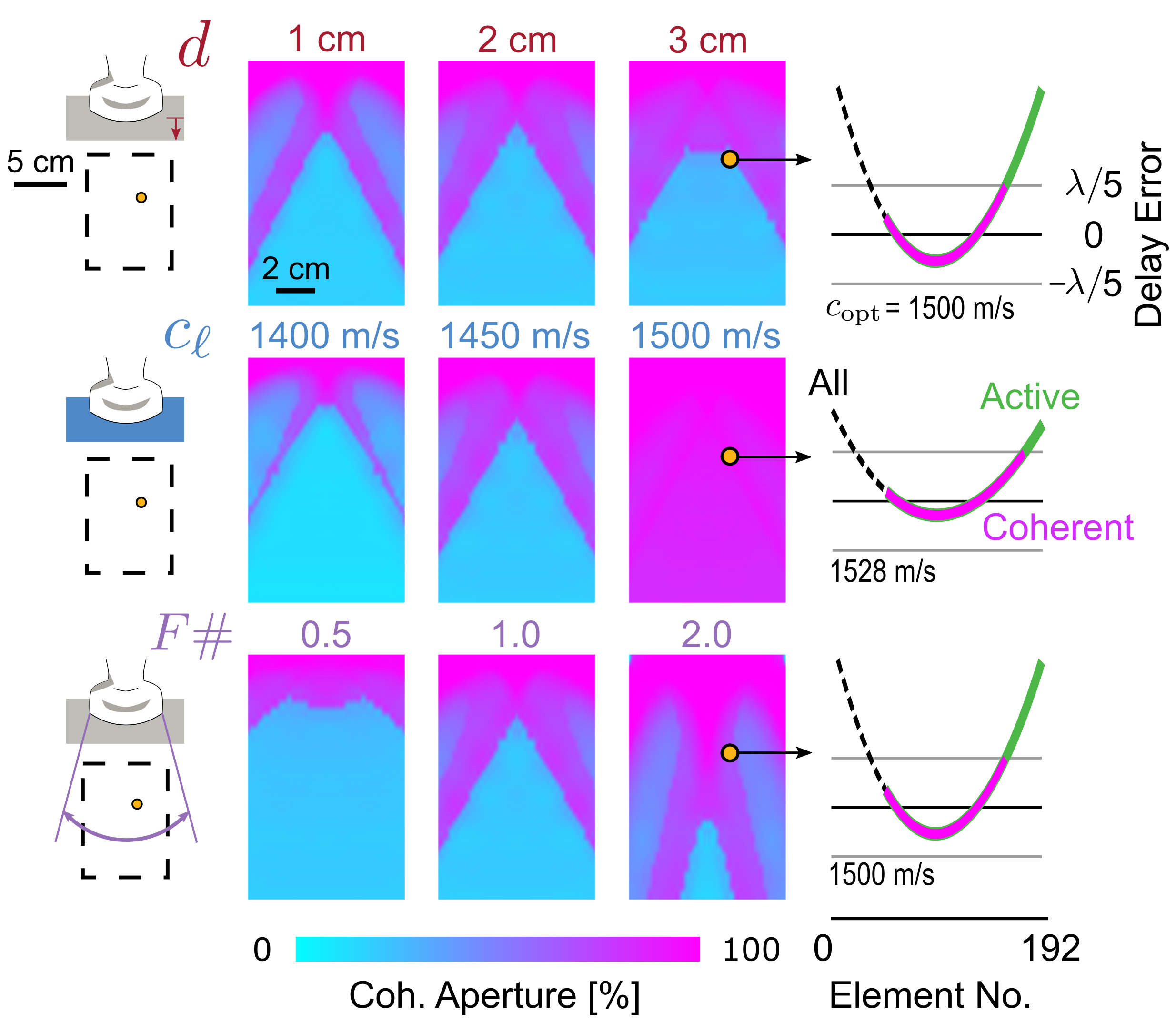}
    \caption{%
      Spatial variation in the coherent aperture for $c_{\mathrm{bf}} = c_{\mathrm{opt}}$ in the region indicated at left, for the indicated layer thicknesses, layer SoS $c_{\ell}$, and receiving aperture for the indicated $F\#$.
      The center column represents the default values for each parameter.
      At right is the arrival time error for each element at the identified $c_{\mathrm{opt}}$ for the location marked by the orange dot.
      The active aperture is shown in green, compared with the coherent fraction (from the colorscale).
      The wavelength (period) is computed for \SI{3}{MHz}.
      }
    \label{fig:ResultsAperture}
\end{figure}

Finally, \cref{fig:ResultsAperture} illustrates the effect of layer properties and $F\#$ on the fraction of the active receive aperture with sufficiently small timing error.
In general, the layer SoS $c_{\ell}$ had the most pronounced effect on the coherent aperture fraction:
Slower layers impart more rapidly changing delay profiles across the face of the array (and thus $c_{\mathrm{opt}}$ varies rapidly with position), and thus $c_{\mathrm{opt}}$ is valid over smaller portions of the active aperture.
However, as seen in \cref{fig:ResultsError} points nearer the array are more likely to have a $c_{\mathrm{opt}}$ that achieves a large coherent aperture fraction.
If a larger $F\#$ are used, the total delay error needs to be optimized over smaller number of elements, and therefore the fraction of these elements that are coherent at $c_{\mathrm{opt}}$ is increased; thus in general, smaller $F\#$ are more likely to supply a sufficiently useful $c_{\mathrm{opt}}$.

\subsection*{Image Metrics}
\begin{figure}
    \centering
    \includegraphics[width=0.9\textwidth]{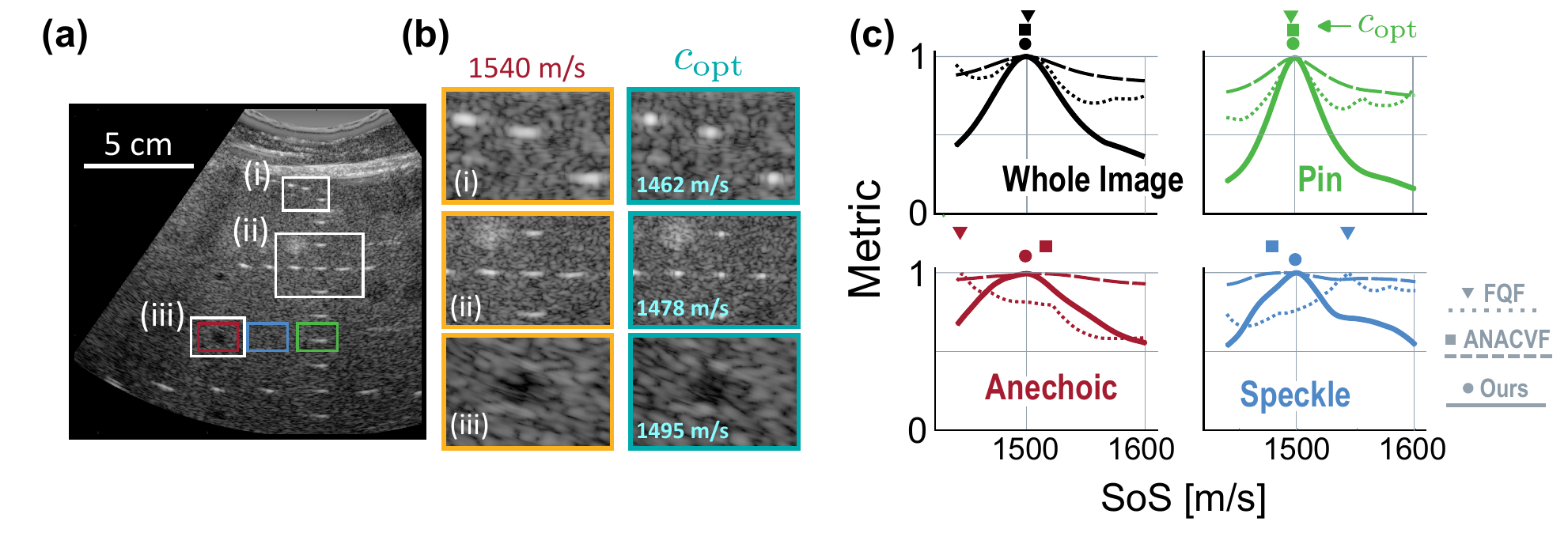}
    \caption{%
      \textbf{(a)}~Example image for a phantom imaged with a \SI{3.5}{cm} fat layer.
      \textbf{(b)}~Enlargement of each ROI in [white boxes in (a)], compared with the image formed at the optimal SoS computed for that ROI and indicated.
      \textbf{(c)}~Comparison of the image metrics calculated for each colored ROI in (a) with the indicated method: 
      focus quality factor~(FQF) from Ref.~\citenum{yoon_optimal_2012} and average normalized covariance function~(ANCVF) from Ref.~\citenum{qu_average_2012}. 
      The markers represent the optimal speed of sound identified by each metric for the indicated ROI.}
    \label{fig:ExampleResult}
\end{figure}
Qualitatively, image regions beamformed at the metric-identified $c_{\mathrm{opt}}$ were superior to the aberrated image at the default \SI{1540}{m/s}.
Specifically, pin targets were most noticeably improved, while contrast targets varied less with changing $c_{\mathrm{bf}}$ [with some improved demarcation of anechoic region boundaries, as shown in \cref{fig:ExampleResult}(a--b)].
The value of $c_{\mathrm{opt}}$ from \cref{eqn:cOptDefinition} demonstrated trends consistent with analysis: note  the estimated $c_{\mathrm{opt}}$ increased with depth, cf. \cref{fig:ResultsSoS} and \cref{fig:ExampleResult}(b).
Additionally, the identified $c_{\mathrm{opt}}$ and metric trends were observed to be robust to target type, especially for anechoic and speckle targets, compared to analogous image metrics [\cref{fig:ExampleResult}(d)].

\begin{figure}
    \centering
    \includegraphics[width=0.85\textwidth]{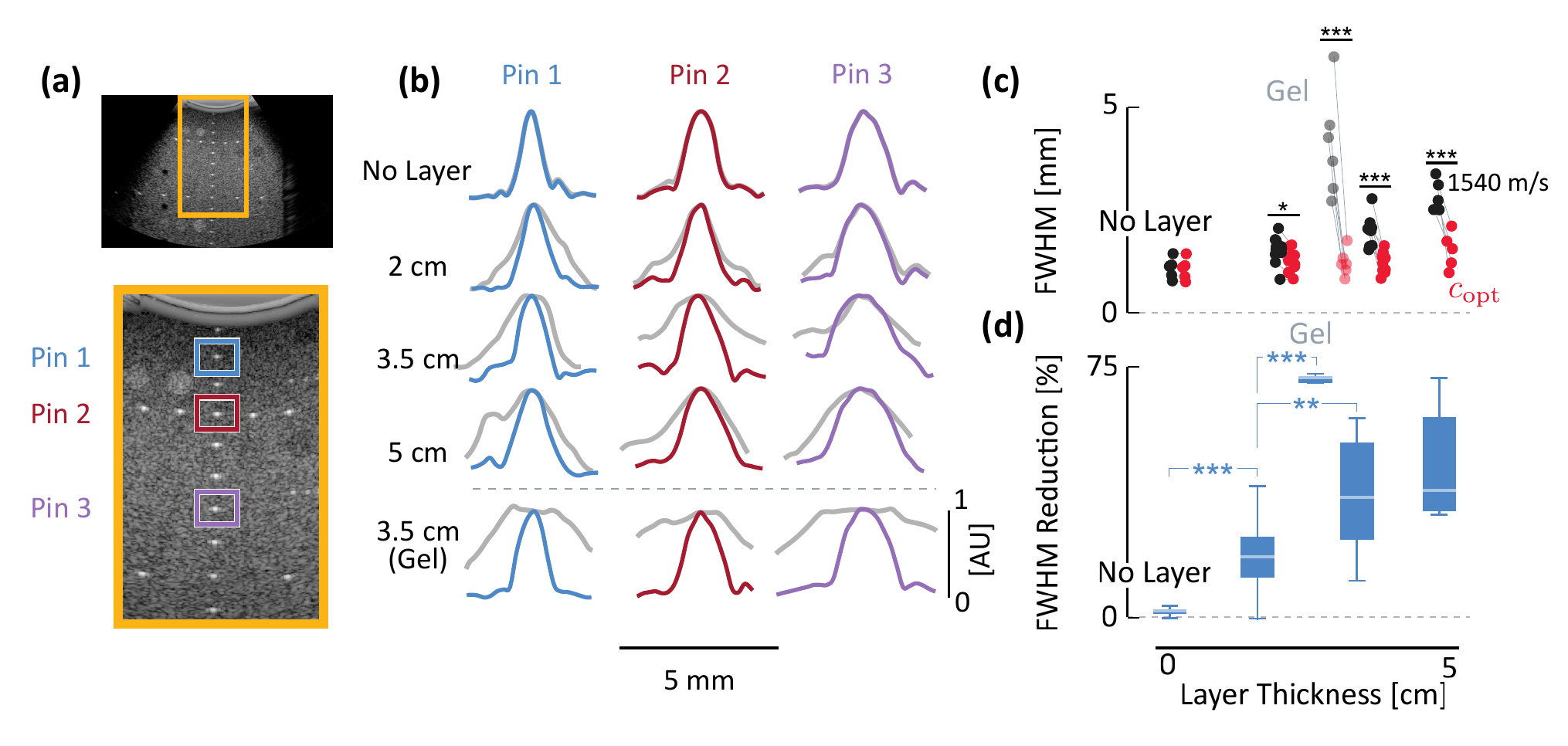}
    \caption{%
      Identification of experimental $c_{\mathrm{opt}}$ with image metrics. 
      \textbf{(a)}~Unaberrated image of the US phantom with inset showing position of pin targets evaluated in (b) for their FWHM.
      \textbf{(b)}~Normalized profiles of the indicated pins from (a). 
      \emph{Color}:~in the image formed with the metric-identified $c_{\mathrm{opt}}$.
       \emph{Gray}:~in the image formed with \SI{1540}{m/s}. 
       Each row corresponds to data from the indicated thickness fat layer.
      \textbf{(c)}~Change in FWHM of pins (6 per image) between the \SI{1540}{m/s} image and the $c_{\mathrm{opt}}$ image. 
      \textbf{(d)}~Relative reduction in the FWHM of the points in (c). 
      In (c) and (d), stars indicate significance as $p$ value of a two-tailed $t$-test; $* = p < 0.05$, $** = p < 0.01$, and ${*}{*}{*} = p < 0.005$}
    \label{fig:ResultsFWHM}
\end{figure}

Over the range of pin resolution targets in the phantom, the identified $c_{\mathrm{opt}}$ reduced the lateral FWHM in the resulting image significantly for all layers compared to the image formed with \SI{1540}{m/s}: from \SI{17 \pm 16}{\percent} for the thinnest 1 cm layer, up to  by \SI{58 \pm 36}{\percent} for a \SI{5}{cm} fat layer.
For the homogeneous gel layer, the improvement was more dramatic (\SI{70 \pm 4}{\percent}), implying that the degree of improvement may be reduced by the specific layer heterogeneity for certain ROIs [\cref{fig:ResultsFWHM}(c--d)].
Generally, the benefit of the correction increased with depth (i.e., deeper targets had greater FWHM reduction relative to the default case), as well as with thicker layers.
These are consistent with the previous analysis, which indicated for a uniform layer, $c_{\mathrm{opt}}$ was most valid for larger values of $d$ (as seen in  \cref{fig:ResultsError}). 
The contrast-to-noise ratio of the resulting images, however, was essentially unchanged between the $c_{\mathrm{opt}}$ and \SI{1540}{m/s} images (difference \SI{0.0 \pm 0.05}{dB}).
While some regions’ CNRs were affected by the layer, individual ROIs maintained comparable contrast before and after correction (see \cref{fig:BulkSoSContrast}).
The small effect on CNR likely due to the much slower variation in contrast with the beamforming SoS and F-number compared to point target FWHM \cite{perrot_so_2021}.

\begin{figure}
    \centering
    \includegraphics[width=0.95\textwidth]{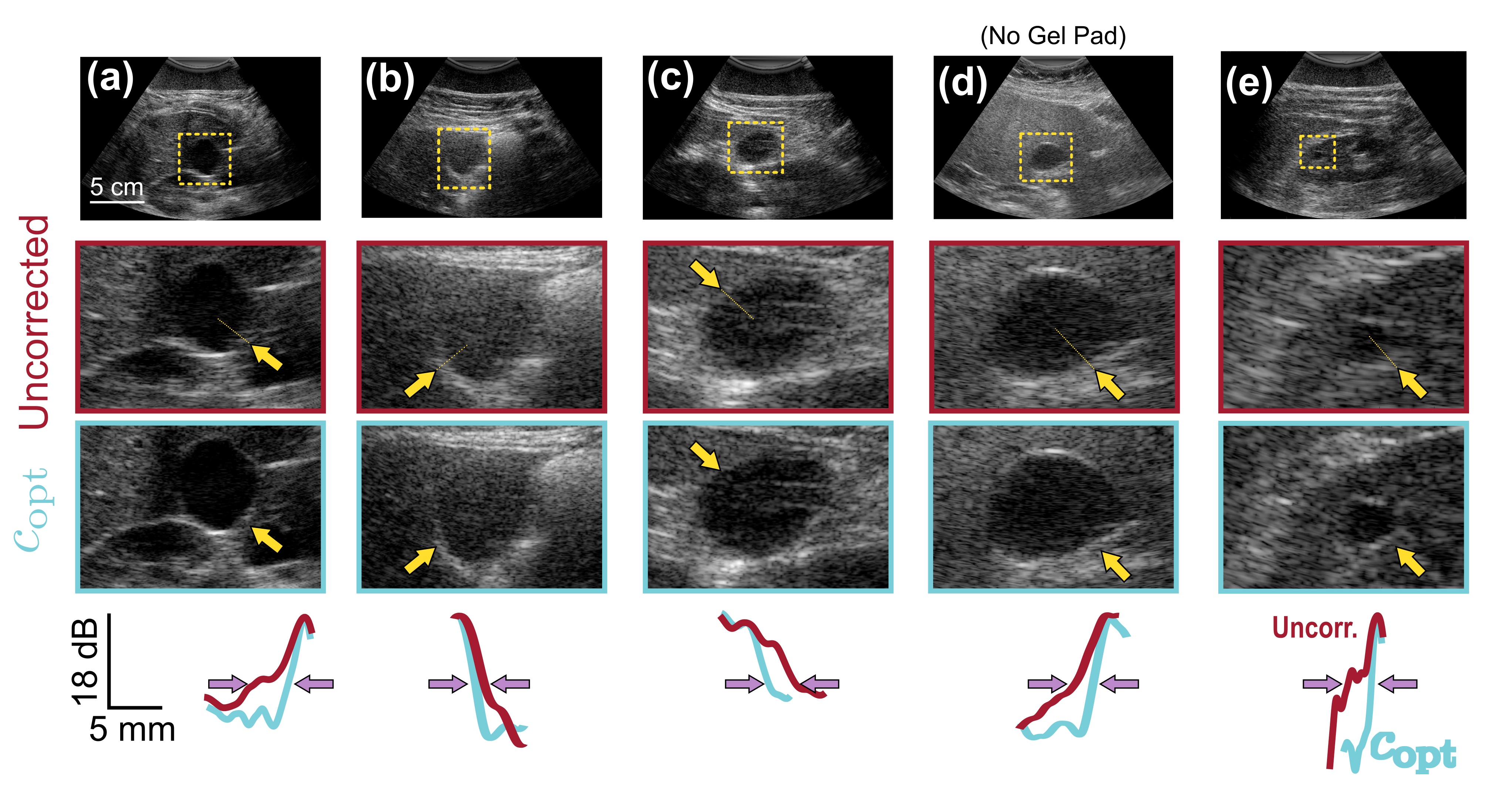}
    \caption{%
     \textbf{(a)}--\textbf{(c)}~Aberrated and corrected images of \textit{in vivo} gallbladders at \SI{3}{MHz}.
     \textbf{(d)}~Gallbladder in subject with suspected NAFLD (no added aberration).
     \textbf{(e)}~Aberrated and corrected images of simple renal cyst.
     Top row: Default images (\SI{1540}{m/s}) with metric ROIs indicated in yellow.
     Second row: ROI of uncorrected image.
     Third row: ROI of image formed with metric identified $c_{\mathrm{opt}}$.
     Note increased sharpness of gallbladder walls (yellow arrows) and reduced clutter within.
     Bottom row: mean intensities from 10 parallel profiles (spacing \SI{0.1}{cm}) at the positions of the yellow arrows; purple arrpws indicate -12~dB difference.
     All image dynamic ranges are \SI{54}{dB}.
      }
    \label{fig:ResultsInVivo}
\end{figure}

\Cref{fig:ResultsInVivo} shows representative \textit{in vivo} imaging results from 5 volunteers. 
In all cases, the aberrator was seen to blur liver structures, particularly near vessel, gallbladder, and cyst boundaries (row 2).
Following identification of the optimal SoS via \cref{eqn:cOptDefinition}, edges were noticeably sharper: the intensity gradient at the indicated walls was larger by a factor of 0.70$\pm$0.78 (bottom row, \cref{fig:ResultsInVivo}).
Consistent with \textit{in vitro} results, the contrast was improved only slightly (per instance improvement of \SI{1.5 \pm 1.5}{\percent}; note difference in gradient, but not necessarily magnitude in the bottom row of \cref{fig:ResultsInVivo}).
Though, this result may be improved, e.g., by weighting the beamformed data with channel coherence mask computed at at $c_{\mathrm{opt}}$, which is expected to provide more reliable contrast enhancement than at 1540 m/s (see \cref{fig:SLSCEffect}).

Finally, we assessed the computational efficiency of the method.
Data beamformed at \SI{10}{m/s} were seen to allow sufficiently accurate interpolation of the images [mean normalized pixel intensity difference $(0.16\pm9.3)\times10^{-3}$ between interpolated and beamformed images].
Thus given DAS beamformer outputs for approximately 10 equally spaced $c_{\mathrm{bf}}$ (e.g., 1460 to \SI{1550}{m/s}), $c_{\mathrm{opt}}$ may be determined to within \SI{1}{m/s} from the image metrics; for the work described herein, this required a few tens of milliseconds for an entire image with minimally optimized \textsc{matlab} code (32~GB RAM, Intel 10 cores at 2.8~GHz).
We also observed (see \cref{fig:MetricCmpTime}) that the computational time to compute \cref{eqn:TotalMetric}: 
(1)~was similar to the FQF (Ref.~\citenum{yoon_optimal_2012}) and substantially less than the ANACVF (Ref.~\citenum{qu_average_2012}) for ROIs larger than $50\times50$ pixels; 
(2)~had only weak dependence on the ROI size; and 
(3)~scaled linearly with the number of trial $c_{\mathrm{bf}}$.
As the multiple DAS operations require several repeated interpolations of the data, overall efficiency could also be improved by performing the beamforming with only the ROI to identify $c_{\mathrm opt}$.

\section{Discussion}
\label{sec:Discussion}
In this work, we established the feasibility and applicability bounds of a single optimal receive beamforming SoS to achieve appreciably low receive delay errors, and to enhance image quality at scales and geometries relevant to abdominal ultrasound.
Through analytical methods, we found that the optimal SoS varies largely with the axial direction, and that we could achieve per-element error of less than a quarter period for a curved array at \SI{3.0}{MHz}.
Thicker layers, those with SoS similar to the underlying medium, and narrower receive apertures (i.e., larger $F\#$) resulted in the smallest arrival time errors for the optimal SoS compared to the ray calculations, consistent with previous work \cite{anderson_direct_1998} (see \cref{fig:ResultsSoS,fig:ResultsError,fig:ResultsAperture}).
The latter effect suggests a 2D optimization over both $F\#$ and $c_{\mathrm{opt}}$ might yield further improvement.
For \textit{in vitro} data, we found the experimentally identified $c_{\mathrm{opt}}$ reduced the point spread function by on the order of \SI{35}{\percent}, and up to $\SI{75}{\percent}$ for uniform layers (\cref{fig:ExampleResult,fig:ResultsFWHM}) compared to the uncorrected case; increased sharpness of organ walls was observed \textit{in vivo} (\cref{fig:ResultsInVivo}), which may aid diagnosis and morphological distinction in typically difficult-to-image subjects.
Thus identification of an optimal receive SoS with the image-based metric thus represents a simple and effective means to improve abdominal ultrasound images.

The proposed method does have several limitations.
First, the analysis assumes a uniform layer and thus ignores any morphological complexity of the layers, which can have important effects on the uniformity and distribution of the arrival time errors.\cite{hinkelman_effect_1998,mast_effect_1998}
Additionally (as will all bulk corrections), random phase errors or more complicated geometries (e.g., a layer of varying thickness) result in delay profiles which cannot be well approximated by a single SoS curve over a useful aperture.
Furthermore, the scheme considers only manipulation of the receive delays; this which enables its use with focused imaging (as here) or unfocused (e.g., plane wave imaging\cite{montaldo_coherent_2009}) techniques.
Further improvement (and agreement with the analysis) might be seen if adjustment of the transmit pattern were also incorporated.
Finally, while the image metrics proved agnostic to ROI size and target type (\cref{fig:MethodsMetrics,fig:ExampleResult}), it is conceivable that imaging parameters held constant in this study (e.g., time gain compensation) may influence their behavior.

We note additionally that as a means to determine an optimal bulk SoS, this technique also may be used as a starting point for other approaches.
For instance, minimum variance beamforming\cite{synnevag_benefits_2009,chau_effects_2018} or coherence-based techniques\cite{lediju_short-lag_2011,nair_robust_2018,long_spatial_2022} may provide greater benefit with initial data delayed at $c_{\mathrm{opt}}$ (see \cref{fig:SLSCEffect}).
Methods for identification of phase shifts from channel correlations\cite{gauss_wavefront_2001,krishnan_adaptive_1997,rigby_beamforming_1995} or matrix imaging techniques\cite{bendjador_svd_2020,lambert_ultrasound_2021} also would be aided, since the expected magnitude of the delay corrections would be smaller relative to those found from \SI{1540}{m/s}.
Finally, recent work has demonstrated that estimation of channel domain data may be obtained from spectral filtering of the beamformer output,\cite{lou_k-space-based_2020,shin_k-space_2022} suggesting that more flexible channel based aberration correction could be applied subsequently to such optimally delayed data without the associated increase in dataset size.

\section{Conclusions}
\label{sec:Conclusion}
In this work we have demonstrated that a single, optimal receive beamforming SoS may provide significant improvement over centimeter scales in layer-aberrated US images.
Specifically, for layered geometries at frequencies relevant to abdominal US, mean arrival time errors less than a fifth of a wavelength were attainable.
In experiments with layered abberators, image metrics (which use only conventional DAS beamformer outputs) reliably identified an optimal SoS for ROIs, improved point target resolution by up to \SI{60}{\percent} \textit{in vitro}, and improved boundary delineation \textit{in vivo}.
Together, these results indicate that identification of $c_{\mathrm{opt}}$ may have immediate application to clinical application and allow improved abdominal imaging among populations for whom it is most crucial.

\section*{Acknowledgments}
The authors acknowledge valuable discussions with Wayne Rigby and Ernest Feleppa in the preparation of this manuscript, as well as the many thorough and helpful insights provided by the anonymous reviewers.
Work funded by GE Healthcare.

\section*{Appendix}
\appendix

\setcounter{equation}{0}
\setcounter{figure}{0}
\renewcommand{\thetable}{A-\arabic{table}}
\renewcommand{\thefigure}{A-\arabic{figure}}
\renewcommand\theequation{A-\arabic{equation}}

\section{Sharpness Metric}
Reference~\citenum{zhu_no-reference_2009} proposes the image sharpness metric 
\begin{align}
    M_{S} = \frac{\sqrt{s_{1}^{2} + \xi N^{2}\sigma^{2}}}{\epsilon + \sigma^{2}}\,,
\end{align}
where $N^{2}$ is the number of pixels in the ROI, $\sigma^{2}$ is the variance of the zero mean Gaussian white noise, $\xi \sim 1/2$ is a scaling factor that depends on the nature of the calculation of the numerical gradients, and $\epsilon$ is a small positive constant to prevent a vanishing denominator.
Here $s_{1}$ is the first singular value of the matrix
\begin{align}
    \begin{pmatrix}
      \big| & \big| \\
      \partial I/\partial x & \partial I/\partial z \\
       \big| & \big|
    \end{pmatrix}\,,
\end{align}
where $I$ is the pixel intensity.
We may approximate
\begin{align}
    M_{S}
    &=
    s_{1}\frac{\left[ 1 + \xi(N\sigma)^{2}/s_{1}^{2}\right]^{\frac{1}{2}}}{\epsilon + \sigma^{2}} 
    \nonumber \\
    &\simeq 
    s_{1}\left[ \frac{1 + (N\sigma)^{2}/2s_{1}^{2}}{\epsilon + \sigma^{2}} \right]\,.
\end{align}
where the factor of $\xi$ has been dropped since we are interested in the order of the second term in the numerator.
If we assume that we have small variance in the signal compared to the magnitude of the first singular value, then, this reduces to $M_{s} = s_{1}/(\epsilon + \sigma^{2}) \approx s_{1}/\epsilon$.
Therefore, it is safe to use the first singular value as the metric, provided the noise and ROI are sufficiently small. 
See also Ref.~\citenum{benjamin_surgery_2018}.

\section{Transverse and Elevational Delays}
\label{sec:DelayEffects}
\begin{figure}[!ht]
    \centering
    \includegraphics[width=0.32\textwidth]{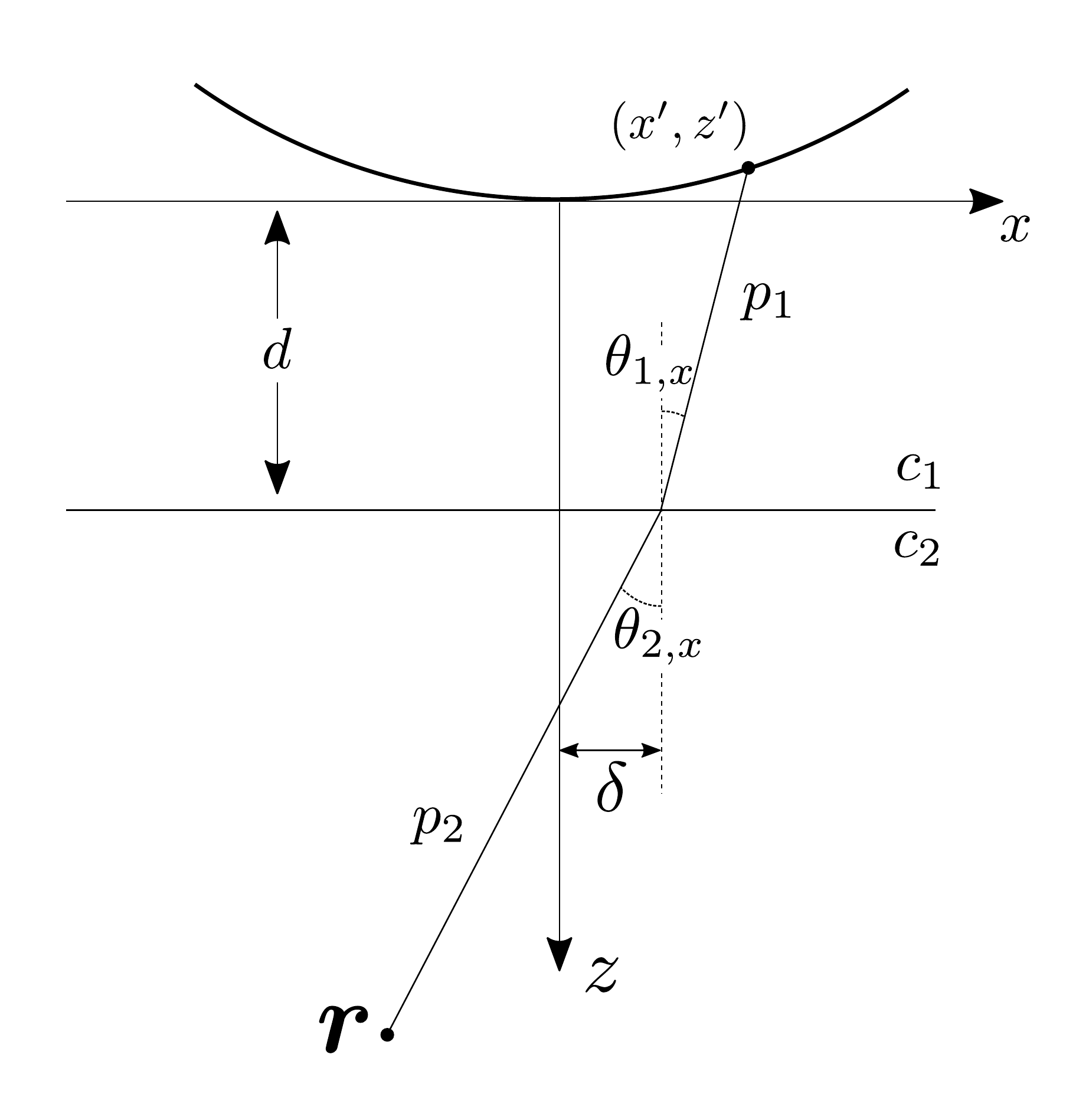}
    \caption{Geometry for solving for the ray.}
    \label{fig:SnellsLawGeometry}
\end{figure}

Call the positions of the individual transducer elements $\boldsymbol{r}_{i}'$, and the positional coordinate is $\boldsymbol{r}$.
The vector that points from $\boldsymbol{r}$ to $\boldsymbol{r}'$ will be defined $\brc = \boldsymbol{r} - \boldsymbol{r}'$.
The coordinate system is not very important, except that the interface between the layer and the capsule is parallel to the $x$-$y$ plane.
For convenience we'll say the reference transducer element (e.g., the center element) is at the origin.
See \cref{fig:SnellsLawGeometry}

Now, consider a ray originating at $\boldsymbol{r}$ and arriving at $\boldsymbol{r}'$, as viewed looking along the $y$-axis (i.e., the visible space is parallel to the $x$-$z$ plane); see \cref{fig:SnellsLawGeometry}.
The ray will be refracted at the interface according to Snell's law
\begin{align}
  \frac{\sin{\theta_{1,x}}}{c_{1}}
  &=
  \frac{\sin{\theta_{2,x}}}{c_{2}}\,,
    \label{eqn:XSnellsLaw}
\end{align}
where $c_{1}$ is the speed of sound in the layer, and $c_{2}$ in the liver.
Now because the $x$-distance between the two points $\boldsymbol{r}$ and $\boldsymbol{r}'$ is known, we have
\begin{align}
    x - x'
    &=
    d\tan{\theta_{1,x}}
    +
    (z - d)\tan{\theta_{2,x}}\,.
    \label{eqn:XDistance}
\end{align}
Now, \cref{eqn:XSnellsLaw} and \cref{eqn:XDistance} form a system of equations of two equations for two unknowns, which may be solved for $\theta_{x,1}$ and $\theta_{x,2}$.

By the same reasoning, we have for the $y$-direction
\begin{align}
    y - y'
    &=
    d\tan{\theta_{1,y}}
    +
    (z - d)\tan{\theta_{2,y}}\,.
    \label{eqn:YDistance}
\end{align}
subject to
\begin{align}
  \frac{\sin{\theta_{1,y}}}{c_{1}}
  &=
  \frac{\sin{\theta_{2,y}}}{c_{2}}\,,
    \label{eqn:YSnellsLaw}
\end{align}

Once we have determined $\theta_{x, 1}$, $\theta_{x, 2}$, $\theta_{y, 1}$, $\theta_{y, 2}$ from \cref{eqn:XSnellsLaw,eqn:XDistance,eqn:YSnellsLaw,eqn:YDistance}, we can compute the travel time $\tau_{l}$ of the ray along the path from $\boldsymbol{r}$ to $\boldsymbol{r}'$.
The travel time in the layer is
\begin{align}
    \tau_{1}
    &=
    \sqrt{%
    d^{2} + %
    \left[d\tan{\theta_{x,1}}\right]^{2} + %
    \left[d\tan{\theta_{y,1}}\right]^{2}
    }%
    \bigg/c_{1}
    \nonumber \\
    &=
    d\sqrt{%
    1 + %
    \tan^{2}{\theta_{x,1}} + %
    \tan^{2}{\theta_{y,1}}
    }%
    \bigg/c_{1}\,.
    \label{eqn:travelTimeLayer}
\end{align}
The travel time in the liver
\begin{align}
    \tau_{2}
    &=
    \sqrt{%
    (z - d)^{2} + %
    \left[(z - d)\tan{\theta_{x,2}}\right]^{2} + %
    \left[(z - d)\tan{\theta_{y,2}}\right]^{2}
    }%
    \bigg/c_{2}
    \nonumber \\
    &=
    (z - d)\sqrt{%
    1 + %
    \tan^{2}{\theta_{x,2}} + %
    \tan^{2}{\theta_{y,2}}
    }%
    \bigg/c_{2}\,.
\end{align}
Now in the no layer case, the travel time is simply
\begin{align}
    \tau_{0} 
    &= 
    \| \boldsymbol{r} - \boldsymbol{r}' \|\Big/c_{2}
    \nonumber \\
    &=
    \rc/c_{2}\,.
\end{align}
Thus, the additional time delay is
\begin{align}
  \tau_{l} = \tau_{1} + \tau_{2} - \tau_{0}.
  \label{eqn:TimeDifference}
\end{align}
In the case of a curved array, the $d$ in \cref{eqn:XDistance,eqn:YDistance,eqn:travelTimeLayer} is replaced by $d \rightarrow d - z'$.
\begin{figure}
    \centering
    \includegraphics[width=0.6\textwidth]{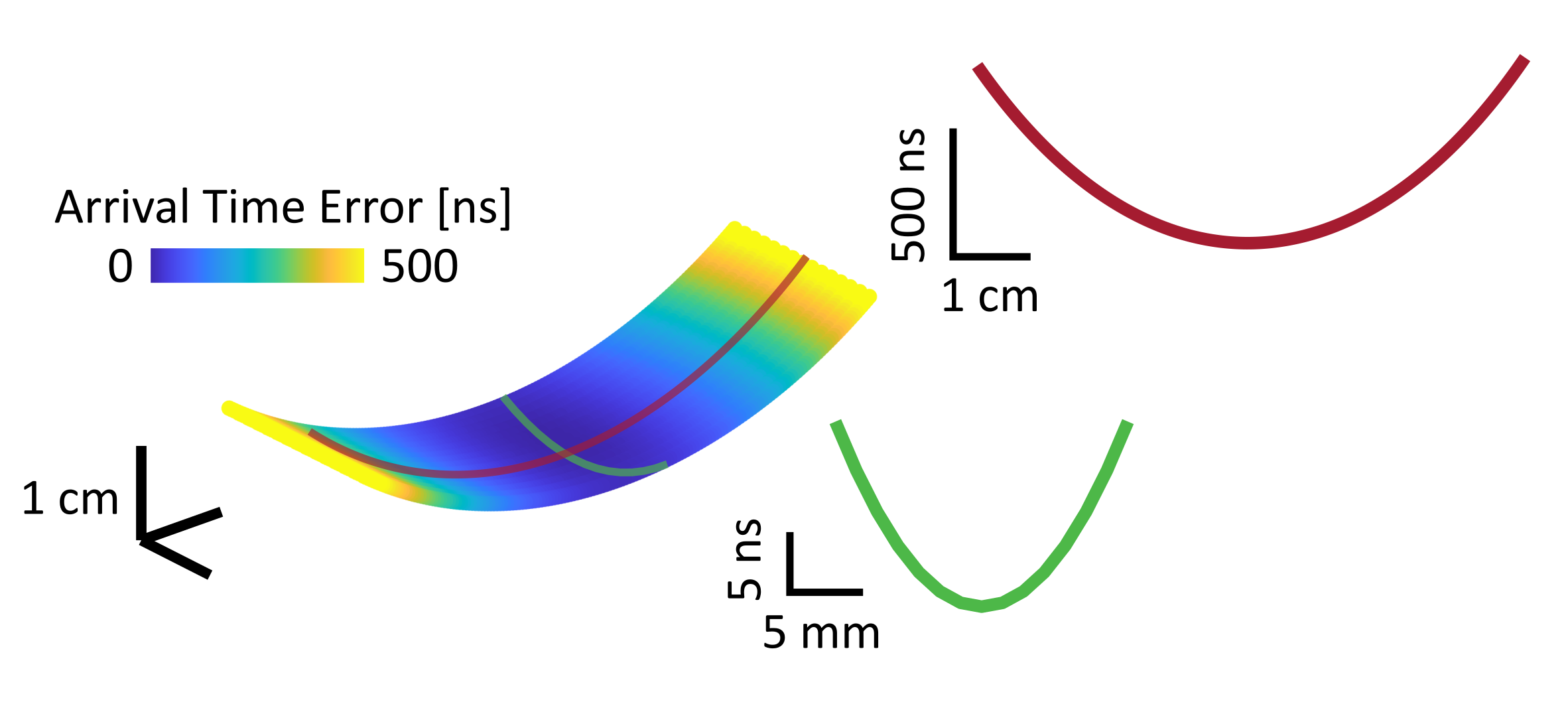}
    \caption{Variation in arrival time error in the transverse (red) and elevational (green) directions;
    note the hundredfold difference in vertical scale.
    Calculations are from \cref{eqn:TimeDifference} for a \SI{2}{cm} layer with speed of sound $c_{1} = \SI{1450}{m/s}$.}
    \label{fig:ArrivalTimeErrors}
\end{figure}
A result for a \SI{2}{cm} layer with a curvilinear array is shown in \cref{fig:ArrivalTimeErrors}.
Note the more than tenfold difference in the gradient between the two, suggesting the transverse (i.e., element-direction) variation has the largest effect.

\section{Effect of ROI and Bandwidth on Metrics}
In general, the heuristic component and composite metrics given by Eqs.~(4--8) were robust to target type and ROI size.
While the trend of a metric may be obscured by portions structures that move into or out of the ROI as $c_{\mathrm{bf}}$ varies (e.g., a pin target's spread as $c_{\mathrm{bf}}$ is far from $c_{\mathrm{opt}}$), they were generally restored with an ROI that comprised the structures of interest.
Additionally, the high pass filter metric is similar to that used in Ref.~\citenum{napolitano_sound_2006}, and requires a choice of spatial bandwidth.
However, that work considered only the 1D lateral transform (i.e., function of $k_{x}$ only) at each depth, and thus energy in particular bands is associated with intensity variation only in the transverse direction.
The high pass filter metric considers energy in a 2D spatial bandwidth, which may correspond to sharp transitions in any orientation. This is especially important, e.g., for vessel walls.

\begin{figure}[!h]
    \centering
    \includegraphics[width=0.5\textwidth]{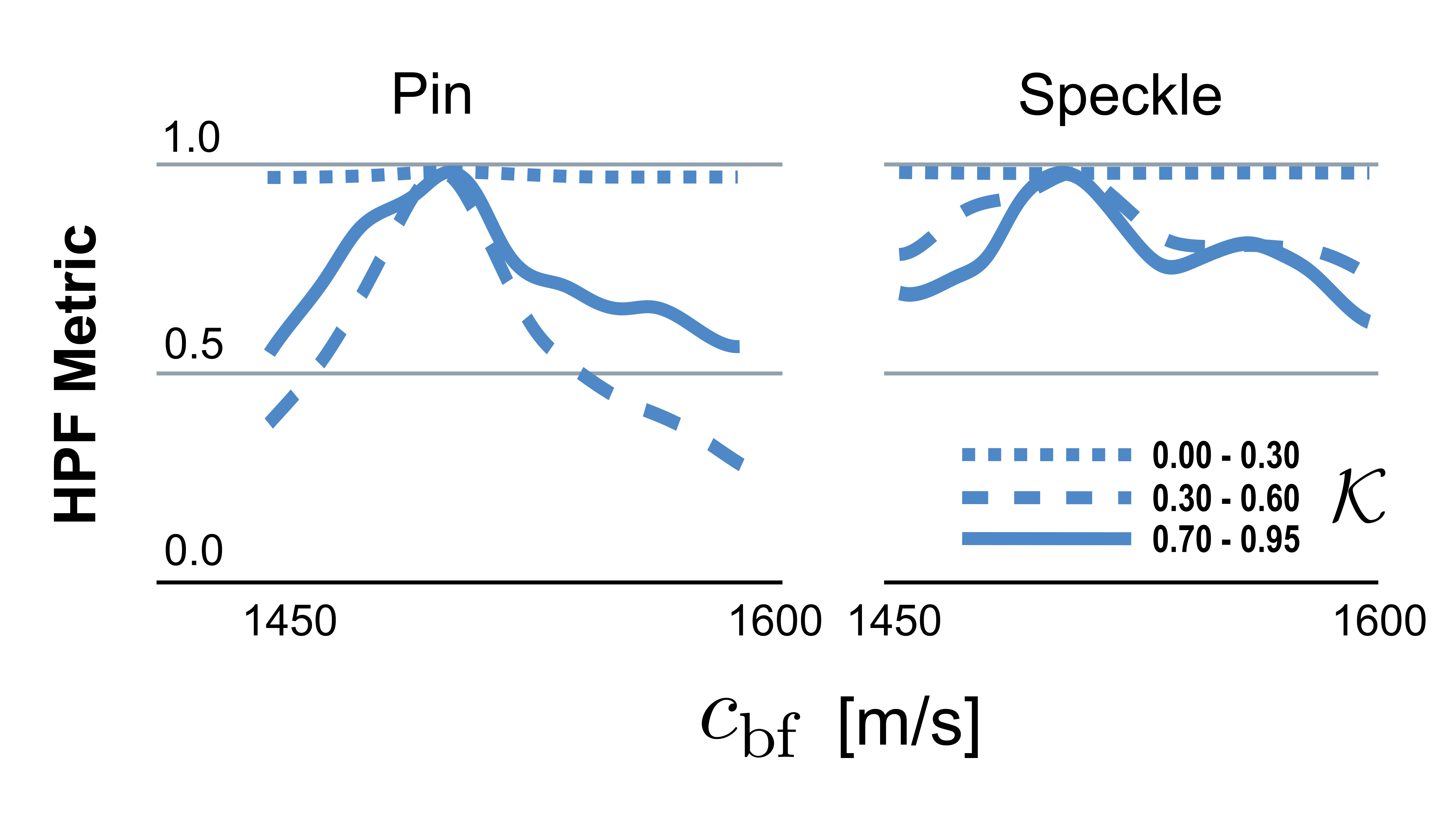}
    \caption{%
    Effect of bandwidth on the value of the high-pass filter metric $M_{HP}$
    Data are for the CIRS phantom with 5~cm fat layer at 3~MHz.%
    }
    \label{fig:HPFMetricBandwidths}
\end{figure}

\Cref{fig:HPFMetricBandwidths} shows the effect of the chosen bandwidth $\mathcal{K}$ on the resulting metric.
Using lower spatial frequencies results in a metric that varies little with $c_[\mathrm{bf}]$, as large-scale features of the image are relatively constant.
However, for higher bandwidths, features to be resolved posses more variable spectral content as they move in and out of focus, resulting in clearer trends in the metric and more reliable identification of $c_{\mathrm{opt}}$.

\section{Effect of Optimal SoS on Contrast and Coherence}
\begin{figure}[!h]
    \centering
    \includegraphics[width=0.65\textwidth]{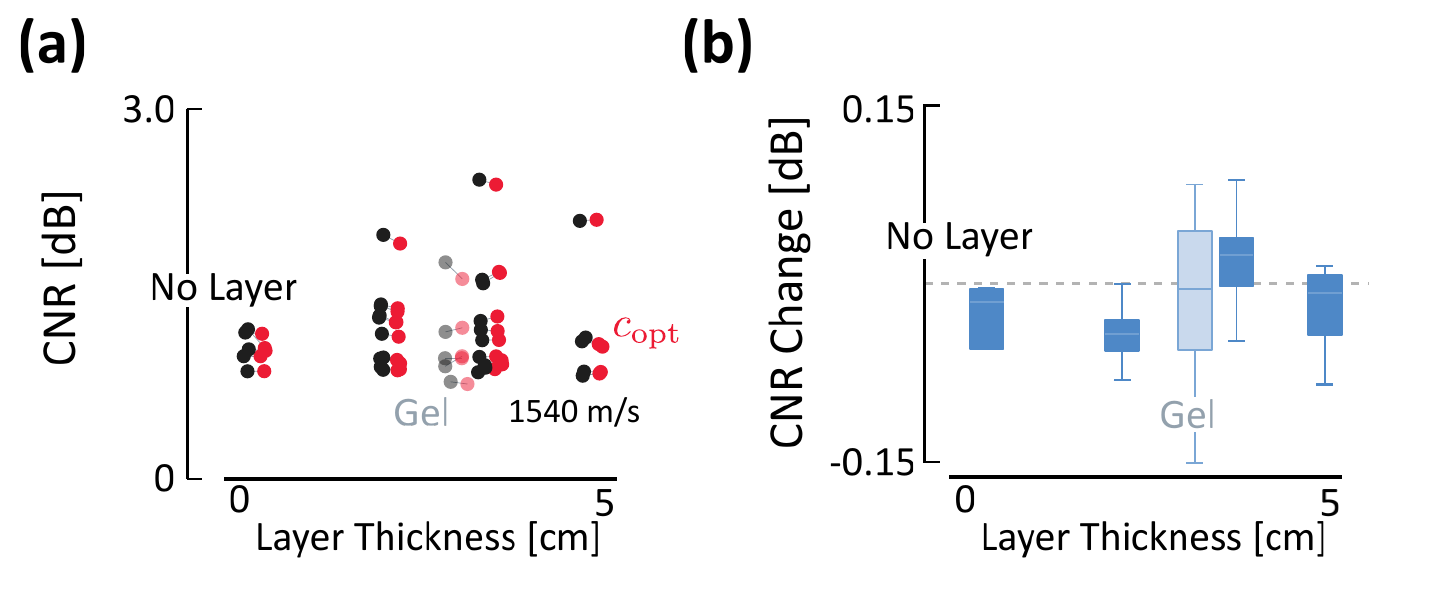}
    \caption{%
    Effect of layer thickness on contrast-to-noise ratio (CNR). 
    \textbf{(a)}~CNR values before (black dots) and after (red dots) bulk SoS correction for 6 contrast targets in the CIRS phantom.
    \textbf{(b)}~Change in the CNR value before and after correction.%
    }
    \label{fig:BulkSoSContrast}
\end{figure}

It was noted that the contrast-to-noise ratio~(CNR) of $N = 5$ lesion-type targets within the phantom were relatively constant between the corrected and uncorrected cases.
\Cref{fig:BulkSoSContrast}(a) shows that while the CNR was slightly affected by the addition of the layers for some targets, the magnitude of these changes were  small.
Additionally, the effect of the beamforming speed of sound $c_{\mathrm{bf}}$ on the contrast was small, and thus even the optimal SoS yielded non-significant changes in the CNR [\cref{fig:BulkSoSContrast}(b)].

\begin{figure}[!ht]
    \centering
    \includegraphics[width=0.55\textwidth]{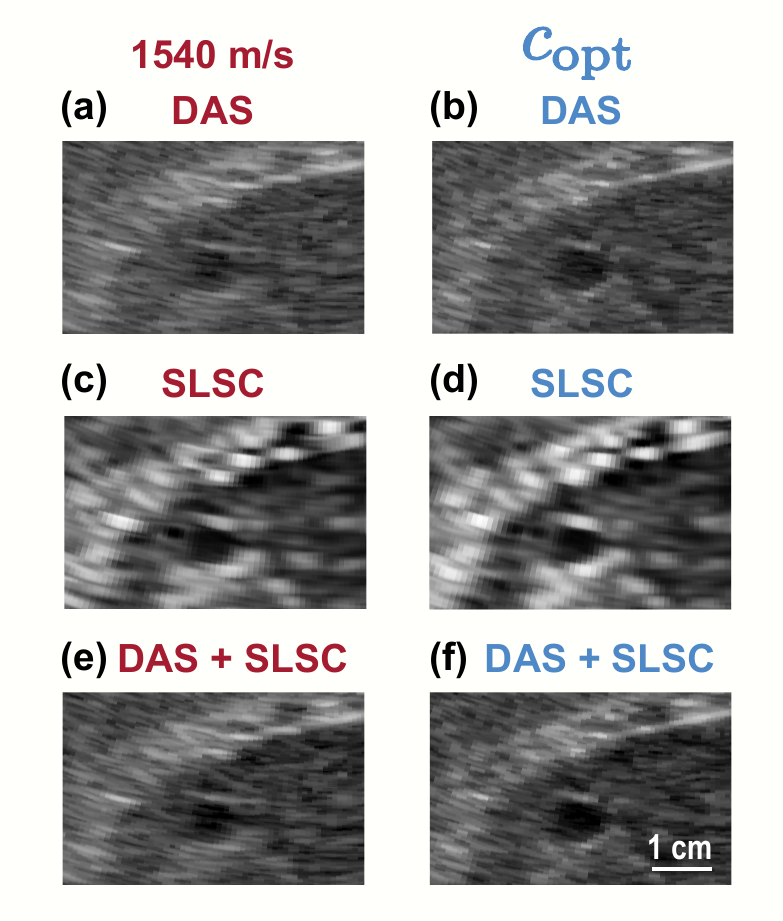}
    \caption{%
    Effect of optimal SoS on signal coherence. %
    \textbf{(a)}~Delay and sum of a simple renal cyst under the aberrating gel layer at 3.0~MHz at 1540~m/s and 
    \textbf{(b)}~at the metric identified optimal SoS of 1450~m/s.
    \textbf{(c)}~Short-lag spatial coherence image (12 lags, kernel size 12 samples) at 1540~m/s and 
    \textbf{(d)}~at $c_{\mathrm{opt}}$.
    \textbf{(e)}~Product of the DAS and SLSC images in (a) and (c), with the SLS mask exponentially weighted by 0.5 [i.e., $I = (I_{\mathrm{SLSC}})^{1/2} \cdot I_{\mathrm{DAS}}$].
    \textbf{(f)}~Same as (e), but with weighted product of (b) and (d).    
    }
    \label{fig:SLSCEffect}
\end{figure}

In addition to improving the image quality of conventional delay and sum images, the metric-identified optimal SoS should also yield improved channel coherence.
That is, delays computed with $c_{\mathrm{opt}}$ should yield better correlation between coherently reflected signals.
To illustrate, \cref{fig:SLSCEffect} shows images formed with default 1540~m/s, and with the metric identified optimal SoS for simple renal cyst under aberrated conditions.
From \cref{fig:SLSCEffect}(a--b), the improvement seen in Fig.~8(e) is replicated.
Additionally \cref{fig:SLSCEffect}(c--d), it is observed that the resulting short-lag spatial coherence~(SLSC) image \cite{hyun_efficient_2017,lediju_short-lag_2011} also appears sharper.
When combining the SLSC image (as an intensity weight to reduce incoherent contributions) with the DAS image [\cref{fig:SLSCEffect}(e--f)], further enhancement of the image is observed.
Such an approach may have value in suppressing signals whose apparent amplitude (even after receive delay correction) is large, but whose coherence with adjacent signals is low (suggesting clutter or aberrant amplitude).

\section{Computational Efficiency}
\begin{figure}[!ht]
    \centering
    \includegraphics[width=0.85\textwidth]{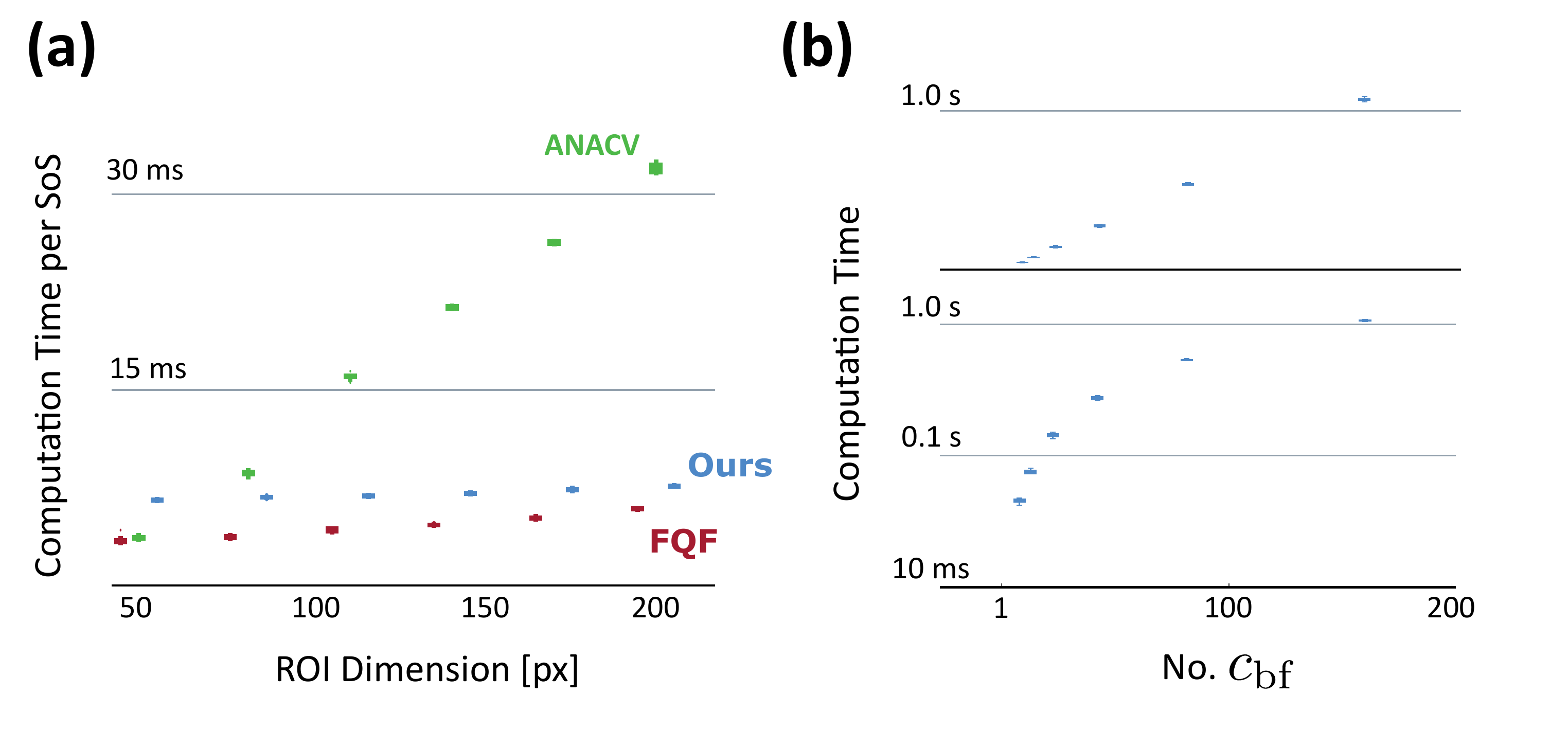}
    \caption{%
    \textbf{(a)}~Computational time for each approach (per trial SoS) as a function of the region of interest dimension (i.e., Dimension 50 $\implies 50^2$ = 2500 pixels). 
    Focus quality factor~(FQF) is from Ref.~\citenum{yoon_optimal_2012}, and average normalized covariance function~(ANACVF) is from Ref.~\citenum{qu_average_2012} with a maximum shift of 6. %
    \textbf{(b)}~Computational time for our metric on a linear (top) and logarithmic (bottom) scale, as a function of the number of candidate beamforming SoSs for an ROI size of 50$\times$50 pixels.
    For both, distributions represent results for 10 ROIs were randomly selected across the beamformed image.
    }
    \label{fig:MetricCmpTime}
\end{figure}

A benefit of the proposed scheme is that it analyses an image dataset that is compressed by a factor on the order of the number of elements compared to methods that compute measures on the channel data.
\Cref{fig:MetricCmpTime}(a) plots the computational efficiency of the method, compared with comparable techniques such as the focusing quality factor \cite{yoon_optimal_2012} and average normalized covariance function \cite{qu_average_2012}.
Our method compares favorably to the ANACV, especially as the size of the ROI increases, and performs similarly to the FQF.
Times are reported for non-optimized versions of each algorithm in MATLAB (2022b) with a general-purpose workstation (Intel Core i9-10900, 32 GB memory).
With respect to the absolute time required, the metric values for tens of speeds of sound could be computed in a few milliseconds for a 2500 pixel region [\cref{fig:MetricCmpTime}(a)], though \cref{fig:MetricCmpTime}(a) suggests that this time does not depend strongly on the ROI size.

\bibliographystyle{unsrt}  
\bibliography{references,references2}  

\end{document}